\documentclass[12pt]{article}
\usepackage[dvips]{graphicx}

\renewcommand{\bar}[1]{\overline{#1}}
\newcommand{\half}{{$\frac{1}{2}$}}

\begin{document}

\begin{flushright}
SLAC-PUB-9271 \\
USM-TH-126\\

\end{flushright}

\bigskip\bigskip
\begin{center}
{\Large \bf Initial-State Interactions  and Single-Spin\\[1ex]
Asymmetries in Drell-Yan Processes} \footnote{\baselineskip=13pt
Work partially supported by the Department of Energy, contract
DE--AC03--76SF00515, by the LG Yonam Foundation, by Fondecyt
(Chile) under grant 8000017 and by MECESUP (Chile) program
FSM9901.}
\end{center}

\vspace{13pt}

\centerline{ \bf Stanley J. Brodsky$^a$, Dae Sung Hwang$^{a,b}$,
and Ivan Schmidt$^{c}$}

\vspace{8pt} {\centerline{$^a$Stanford Linear Accelerator
Center,}}

{\centerline{Stanford University, Stanford, California 94309,
USA}}

\centerline{e-mail: sjbth@slac.stanford.edu}

\vspace{8pt} {\centerline{$^{b}$ Department of Physics, Sejong
University, Seoul 143--747, Korea}}

\centerline{e-mail: dshwang@sejong.ac.kr}

\vspace{8pt} {\centerline {$^{c}$Departamento de F\'\i sica,
Universidad T\'ecnica Federico Santa Mar\'\i a,}}

{\centerline {Casilla 110-V, 
Valpara\'\i so, Chile}}

\centerline{e-mail: ischmidt@fis.utfsm.cl }

\vfill

\centerline{Submitted to Nuclear Physics B.}
\vfill

\newpage

\setlength{\baselineskip}{13pt}

\bigskip

\begin{abstract}
We show that the initial-state interactions from gluon exchange
between the incoming quark and the target spectator system lead to
leading-twist single-spin asymmetries in the Drell-Yan process
$H_1 H_2^\uparrow \to \ell^+ \ell^- X$.  The QCD initial-state
interactions produce a $T-$odd spin-correlation $\vec S_{H_2}\cdot
\vec P_{H_1} \times \vec Q$ between the target spin and the
virtual photon production plane which is not power-law suppressed
in the Drell-Yan scaling limit at large photon virtuality $Q^2$ at
fixed $x_F$.  The single-spin asymmetry which arises from the
initial-state interactions is not related to the target or
projectile transversity distribution $\delta q_H(x,Q).$  The
origin of the single-spin asymmetry in $\pi p^\uparrow \to \ell^+
\ell^- X$ is a phase difference between two amplitudes coupling
the proton target with $J^z_p = \pm {1\over 2}$ to the same
final-state, the same amplitudes which are necessary to produce a
nonzero proton anomalous magnetic moment.  The calculation
requires the overlap of target light-front wavefunctions with
different orbital angular momentum: $\Delta L^z = 1;$ thus the SSA
in the Drell-Yan reaction provides a direct measure of orbital
angular momentum in the QCD bound state.  The single-spin
asymmetry predicted for the Drell-Yan process $\pi p^\uparrow \to
\ell^+ \ell^- X$ is similar to the single-spin asymmetries in deep
inelastic semi-inclusive leptoproduction $\ell p^\uparrow \to
\ell' \pi X$ which arises from the final-state rescattering of the
outgoing quark.  The Bjorken-scaling single-spin asymmetries
predicted for the Drell-Yan and leptoproduction processes
highlight the importance of initial- and final-state interactions
for QCD observables.

\end{abstract}

 \bigskip \bigskip

\section{Introduction}

Single-spin asymmetries in hadronic reactions provide a remarkable
window to QCD mechanisms at the amplitude level.  In general,
single-spin asymmetries measure the correlation of the spin
projection of a hadron with a production or scattering
plane~\cite{Sivers:1990fh}.  Such correlations are odd under time
reversal, and thus they can arise in a time-reversal
invariant theory only when there is a phase difference between
different spin amplitudes. Specifically, a nonzero correlation of
the proton spin normal to a production plane measures the phase
difference between two amplitudes coupling the proton target with
$J^z_p = \pm {1\over 2}$ to the same final-state.  The calculation
requires the overlap of target light-front wavefunctions with
different orbital angular momentum: $\Delta L^z = 1;$ thus a
single-spin asymmetry (SSA) provides a direct measure of orbital
angular momentum in the QCD bound state.

Consider the SSA produced in semi-inclusive deep inelastic
scattering $\ell p^\uparrow \to \ell^\prime \pi X$.   In the
target rest frame, such a single target spin correlation
corresponds to the $T$-odd triple product $ \vec S_p \cdot \vec
p_\pi \times \vec q.$ (The covariant form of this correlation is
$\epsilon_{\mu \nu \sigma \tau} S^\mu_p p^\nu q^\sigma
p^\tau_\pi.)$   Significant asymmetries $A_{UL}$ and $A_{UT}$ of
this type have in fact been observed for targets polarized
parallel to or transverse to the lepton beam
direction~\cite{hermes0001,smc99}.

In a recent paper~\cite{Brodsky:2002cx} we have  shown that the
QCD final-state interactions (gluon exchange) between the struck
quark and the proton spectator system in semi-inclusive deep
inelastic lepton scattering can produce single-spin asymmetries
which survive in the Bjorken limit.  Such effects are proportional
to the matrix element of a higher-twist quark-quark-gluon
correlator in the target hadron, and thus it has been assumed on
dimensional grounds that any SSA arising from this source must be
suppressed by a power of the momentum transfer $Q$ in the Bjorken
limit. However, another momentum scale enters into the
semi-inclusive process---the transverse momentum ${\vec r}_{\perp}
= {\vec {p_\pi}}_{\perp} - {\vec q}_{\perp}$ of the emitted pion
relative to the photon direction, and we have shown that the
power-law suppression due to the higher-twist quark-quark-gluon
correlator takes the form of an inverse power of $r_\perp$ rather
$Q.$ As shown in the Appendix, $r^2_\perp$ can be written in terms
of the invariant momentum transfer squared $t$ from the proton to
the spectator system and the Bjorken variable.

Corrections from spin-one gluon exchange in the initial- or
final-state of QCD processes are not suppressed at high energies
because the coupling is vector-like. Therefore, as a consequence
of the gauge coupling of QCD, single-spin asymmetries in
semi-inclusive deep inelastic scattering survive in the Bjorken
limit of large $Q^2$ at fixed $x_{bj}$ and fixed $\vec r_\perp$.
Recently it has been shown \cite{Brodsky:2001ue,Peigne:2002iw}
that the same type of final-state interaction is the origin of the
leading-twist diffractive component in deep inelastic scattering,
implying that the pomeron is not a universal property of the
target proton's wavefunction, and that it depends in detail on the
deep inelastic scattering (DIS) process itself. Diffractive
processes in DIS in turn lead to nuclear shadowing in the case of
nuclear targets, showing that shadowing is not an intrinsic
property of nuclear wavefunctions.

The final-state phases which we compute are analogous to the
``Coulomb" phases to the hard subprocess which arises from gauge
interactions between outgoing charge particles in
QED~\cite{Weinberg:1965nx}.  More specifically, we require the
difference  between the gauge interaction phases for the $J^z_p=
\pm {1\over 2}$ amplitudes.  The phases depend on the spin because
the outgoing particles interact at different impact separation
corresponding to their different relative orbital angular
momentum.

In our previous paper~\cite{Brodsky:2002cx}, we explicitly
evaluated the SSA for electroproduction for a specific model of a
spin-\half ~ proton of mass $M$ with charged spin-\half ~ and
spin-0 constituents of mass $m$ and $\lambda$, respectively, as in
the QCD-motivated quark-diquark model of a nucleon.  The basic
leptoproduction reaction is then $\gamma^* p \to q (qq)_0$. Our
analysis predicts a nonzero SSA for the target spin normal to the
photon to quark-jet $ \vec S_p \cdot \vec p_q \times \vec q$ which
can be determined by using a jet variable such as thrust to
determine the current quark direction;  i.e., we predict a SSA
even without final-state jet hadronization.  Our mechanism is thus
distinct from a description of SSA based on transversity and
phased fragmentation functions.

\begin{figure}
\centering
\includegraphics{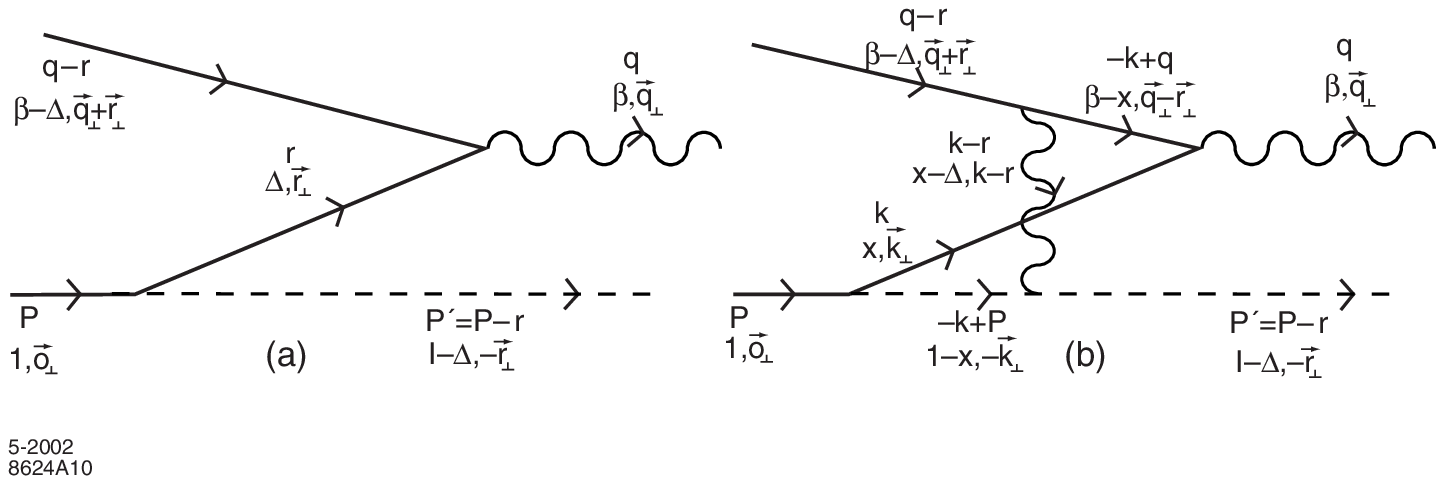}
\caption[*]{The initial-state interaction in the Drell-Yan process.}
\label{fig:1}
\end{figure}

Recently Collins \cite{Collins} has pointed out some important
consequences of these results for SSA in deep inelastic
scattering. In his treatment the final-state interactions of the
struck quark are incorporated into Wilson line path-ordered
exponentials which augment the light-cone wavefunctions. [See
also~\cite {Ji:2002aa}.]  Since the final-state interactions
appear at short light-cone times $\Delta x^+ = {\cal O}(1/\nu)$
after the virtual photon acts, they can be distinguished from
hadronization processes which occur over long times.  Collins has
stressed the fact that single-spin asymmetries probe the partonic
structure associated with chiral-symmetry breaking.  Furthermore,
these results show that time-reversal-odd parton densities are
allowed, opening up a whole range of phenomenological
applications~\cite{Boer:1997nt,Anselmino:1998yz,Boer:1999mm}. In
particular, as noted by Collins, initial-state interactions
between the annihilating antiquark and the spectator system of the
target can produce single-spin asymmetries in the Drell-Yan
process.

In this paper we shall extend our analysis to initial- and
final-state QCD effects to predict single-spin asymmetries in
hadron-induced hard QCD processes.  Specifically, we shall
consider the Drell-Yan (DY) type reactions~\cite{Drell:1970wh}
such as $\pi p^\uparrow \to \ell^+ \ell^- X$.  Here the target
particle is polarized normal to the pion production plane.  The
target spin asymmetry  can be produced due to the initial-state
gluon-exchange interactions between the interacting antiquark
coming from one hadronic system and the spectator system of the
other. This is shown in the diagram of Fig. 1.  The importance of
initial-state interactions in the theory of massive lepton pair
production, $Q_\perp$ broadening, and energy loss in a nuclear
target has been discussed in
Refs.~\cite{Bodwin:1981fv,Bodwin:1988fs,Brodsky:1996nj}.

The orientation of the target spin $S_z = \pm 1/2$ corresponds to
amplitudes differing by relative orbital angular momentum $\Delta
L^z = 1.$ The initial-state interaction from a gluon exchanged
between the annihilating antiquark and target spectator system
depends in detail on this relative orbital angular momentum.  In
contrast, the initial or final-state interactions due to the
exchange of gauge particles between partons not participating in
the hard subprocess do not contribute to the SSA.  Such
spectator-spectator interactions occur at large impact separation
and are not sensitive to just one unit difference $\Delta L^z = 1$
of the orbital angular momentum of the target wavefunction.

Our mechanism thus depends on the interference of different
amplitudes arising from the target hadron's wavefunction and is
distinct from probabilistic measures of the target such as
transversity. It is also important to note that the target spin
asymmetries which we compute in the DY and DIS processes require
the same overlap of wavefunctions which enters the computation of
the target nucleon's magnetic moment.  In addition, by selecting
different initial mesons in the DY process, we can isolate the
flavor of the annihilating quark and antiquark.  The flavor
dependence of single-spin asymmetries thus has the potential to
provide detailed information of the spin and flavor content of
nucleons at the amplitude level.

As in our analysis of semi-inclusive DIS, we shall calculate the
single-spin asymmetry in the Drell-Yan process induced by
initial-state interactions by adopting an effective theory of a
spin-${1\over 2}$ proton of mass $M$ with charged spin-${1\over
2}$ and spin-$0$ constituents of mass $m$ and $\lambda$,
respectively, as in a quark-diquark model. We will take the
initial particle to be just an antiquark.  The result for specific
meson projectiles such as $M p^\uparrow \to \ell^+ \ell^- X$ is
then obtained by convolution with the antiquark distribution of
the incoming meson.  One can also incorporate target nucleon
wavefunctions with a quark-vector diquark structure.  In a more
complete study, one should allow for a many-parton light-front
Fock state wavefunction representation of the target. The results,
however, are always normalized to the quark contribution to the
proton anomalous moment, and thus are basically model-independent.

\section{Crossing}

There is a simple diagrammatic connection between the amplitude
describing the initial-state interaction of the annihilating
antiquark, which gives a single-spin asymmetry for the Drell-Yan
process $\pi p^{\uparrow} \to \ell^+ \ell^- X$, and the
final-state rescattering amplitude of the struck quark, which
gives the single-spin asymmetries in semi-inclusive deep inelastic
leptoproduction $\ell p^{\uparrow}\to \ell' \pi X$. The crossing
of the Feynman amplitude for $\gamma^*(\tilde q) p(P) \to (\tilde
q+r) (P-r)$ in DIS gives $(-\tilde q-r) p(P) \to \gamma^*(-\tilde
q) (P-r)$ for DY by reversing the four-vectors of the photon and
quark lines.  The outgoing quark with momentum $\tilde q+r$ in DIS
becomes the incoming antiquark with momentum $-\tilde q-r$ in DY.
We can use crossing of the Lorentz invariant amplitudes for DIS as
a guide for obtaining the amplitudes for DY
amplitude~\cite{Brandenburg:1994mm}. [In the next section it will
be convenient to label $\tilde q = - q$ with $q^+ > 0.$ ]

In general, one cannot use crossing to relate imaginary parts of
amplitudes to each other, since under crossing, real and imaginary
parts become connected. However, in our case, the relevant
one-gluon exchange diagrams in DIS and DY are both purely
imaginary at high energy, so their magnitudes are related by
crossing. Thus a crucial test of our mechanism is an exact
relation between the magnitude and flavor dependence of the SSA in
the Drell-Yan reaction and the SSA in deep inelastic scattering.
We thus predict the DY SSA of the proton spin with the normal to
the antiquark to virtual photon plane: $\vec S_p \cdot \vec
p_{\bar q} \times \vec{\tilde q}$.  It is identical -- up to a
sign -- to the SSA computed in DIS for $\vec S_p \cdot \vec p_{ q}
\times \vec q$.

The phase arising from the initial- and final-state interactions
in QCD is analogous to the Coulomb phase of Abelian QED
amplitudes. The Coulomb phase depends on the  product of charges
and relative velocity of each ingoing and outgoing charged
pair~\cite{Weinberg:1965nx}.  Thus the sign of the phase in DY and
DIS are opposite because of the different color charge of the
ingoing $\overline 3_C$ antiquark in DY and the outgoing $3_C$
quark in DIS.  In order to check the sign, we will carry out the
DY calculation explicitly in the next section.

The asymmetry in the Drell-Yan process is thus the same as that
obtained in DIS, with the appropriate identification of variables,
but with the opposite sign.  This has been stressed recently by
Collins~\cite{Collins}.  Therefore the single-spin asymmetry
transverse to the production plane in the Drell-Yan process can be
obtained from the results of our recent
paper~\cite{Brodsky:2002cx}:
\begin{eqnarray}
{\cal P}_y &=& -\ {e_1e_2\over 8\pi} \ {2\ \Bigl(\ \Delta\, M+m\
\Bigr)\ r^1\over \Big[\ \Bigl( \ \Delta\, M+m\ \Bigr)^2\ +\ {\vec
r}_{\perp}^2\ \Big]}\ \Big[\ {\vec r}_{\perp}^2+\Delta
(1-\Delta)(-M^2+{m^2\over\Delta} +{\lambda^2\over 1-\Delta})\
\Big] \nonumber\\ &\times&
\
{1\over {\vec r}_{\perp}^2}\ {\rm ln}{{\vec r}_{\perp}^2 +\Delta
(1-\Delta)(-M^2+{m^2\over\Delta}+{\lambda^2\over 1-\Delta})\over
\Delta (1-\Delta)(-M^2+{m^2\over\Delta}+{\lambda^2\over
1-\Delta})}\ . \label{sa2b}
\end{eqnarray}
Here $\Delta  = {q^2\over 2P\cdot q } = {q^2 \over 2 M \nu} $
where $\nu$ is the energy of the lepton pair in the target rest
frame.  The kinematics are given in detail in the Appendix.

\section{Calculation}

We can check the results of the previous section obtained using
crossing by performing a direct calculation of the $\bar q
p^\uparrow \to \gamma^* (qq)_0$ amplitude where we take a
spin-zero diquark for the proton spectator system.  The kinematics
are $(q-r) p(P) \to \gamma^*(q) (P-r),$ as in Fig. 1. The $J^z = +
{1\over 2}$ two-particle Fock state is given by
\cite{BD80,Brodsky:2000ii}
\begin{eqnarray}
&&\left|\Psi^{\uparrow}_{\rm two \ particle}(P^+, \vec P_\perp = \vec
0_\perp)\right>
\label{sn1}\nonumber \\
&=& \int\frac{{\mathrm d}^2 {\vec k}_{\perp} {\mathrm d} x
}{{\sqrt{x(1-x)}}16 \pi^3} \Big[ \ \psi^{\uparrow}_{+\frac{1}{2}}
(x,{\vec k}_{\perp})\, \left| +\frac{1}{2}\, ;\,\, xP^+\, ,\,\,
{\vec k}_{\perp} \right>  \\
&& \qquad +\psi^{\uparrow}_{-\frac{1}{2}} (x,{\vec k}_{\perp})\,
\left| -\frac{1}{2}\, ;\,\, xP^+\, ,\,\, {\vec k}_{\perp} \right>\
\Big]\ , \nonumber
\end{eqnarray}
where
\begin{equation}
\left
\{ \begin{array}{l}
\psi^{\uparrow}_{+\frac{1}{2}} (x,{\vec k}_{\perp})=(M+\frac{m}{x})\,
\varphi \ ,\\
\psi^{\uparrow}_{-\frac{1}{2}} (x,{\vec k}_{\perp})=
-\frac{(+k^1+{\mathrm i} k^2)}{x }\,
\varphi \ .
\end{array}
\right.
\label{sn2}
\end{equation}
The scalar part of the wavefunction $\varphi$ depends on the dynamics.
In the perturbative theory it is simply
\begin{equation}
\varphi=\varphi (x,{\vec k}_{\perp})=\frac{ {e\over \sqrt{1-x}}}{M^2-{{\vec
k}_{\perp}^2+m^2\over x}-{{\vec k}_{\perp}^2+\lambda^2\over 1-x}}\ .
\label{wfdenom}
\end{equation}
In general one normalizes the Fock state to unit probability.

Similarly, the $J^z = - {1\over 2}$ two-particle Fock state has
two components
\begin{equation}
\left
\{ \begin{array}{l}
\psi^{\downarrow}_{+\frac{1}{2}} (x,{\vec k}_{\perp})=
\frac{(+k^1-{\mathrm i} k^2)}{x }\,
\varphi \ ,\\
\psi^{\downarrow}_{-\frac{1}{2}} (x,{\vec k}_{\perp})=(M+\frac{m}{x})\,
\varphi \ .
\end{array}
\right.
\label{sn2a}
\end{equation}
The spin-flip amplitudes in (\ref{sn2}) and (\ref{sn2a}) have
orbital angular momentum projection $l^z =+1$ and $-1$
respectively.  The numerator structure of the wavefunctions is
characteristic of the orbital angular momentum, and holds for both
perturbative and non-perturbative couplings.

We require the interference between the tree amplitude of Fig. 1(a)
and the one-loop amplitude of Fig. 1(b).  The
contributing amplitudes have the
following structure through one-loop order:
\begin{eqnarray}
{\cal A}(\Uparrow \to \downarrow)&=&(M+{m\over \Delta})\ C\
(h+i{e_1e_2\over 8\pi}g_1)
\label{s1}\\
{\cal A}(\Downarrow \to \downarrow)&=&\ ({+r^1-ir^2\over \Delta})\ \ C\
(h+i{e_1e_2\over 8\pi}g_2)
\label{s2}\\
{\cal A}(\Uparrow \to \uparrow)&=&\ ({-r^1-ir^2\over \Delta})\ \ C\
(h+i{e_1e_2\over 8\pi}g_2)
\label{s3}\\
{\cal A}(\Downarrow \to \uparrow)&=&(M+{m\over \Delta})\ C\
(h+i{e_1e_2\over 8\pi}g_1) \ , \label{s4}
\end{eqnarray}
where
\begin{eqnarray}
C&=&-\ g\ e_1\ P^+\ {\sqrt{\beta - \Delta}}\ 2\ \Delta\ (1-\Delta)
\label{s5}\\
h&=& {1\over {\vec r}_{\perp}^2+\Delta
(1-\Delta)(-M^2+{m^2\over\Delta} +{\lambda^2\over 1-\Delta})}\ .
\label{s6}
\end{eqnarray}
The label $\Uparrow/\Downarrow$ corresponds to $J^z_p = \pm
{1\over 2}$ of the proton spin. The second label
$\uparrow/\downarrow$ gives the spin projection $J^z_q = \pm
{1\over 2}$ of the interacting spin-\half~ constituent antiquark
of the other proton. Here $e_1$ and $e_2$ are the electric charges
of the proton's constituents $q$ and $(qq)_0$, respectively, and
$g$ is the coupling constant of the effective proton-$q$-$(qq)_0$
vertex. The first term in (\ref{s1}) to (\ref{s4}) is the Born
contribution of the tree graph.  The crucial result will be the
fact that the contributions $g_1$ and $g_2$ from the one-loop
diagram Fig. 1(b) are different, and that their difference is
infrared finite.  A gauge boson mass $\lambda_g$ will be used as
an infrared regulator in the calculation of $g_1$ and $g_2.$ The
final result for $g_1- g_2$ is infrared finite, and $\lambda_g$
can be set to zero. The calculation will be done using light-cone
time-ordered perturbation theory, or equivalently, by integrating
Feynman loop-diagrams over $dk^-$.

We take $\vec q$ to lie in the $\hat z - \hat x$ plane,
$\vec q = (q^x, q^y, q^z) = (q^1, 0, q^3)$.  We denote $q^+$ as
\begin{equation}
q^+\ =\ \beta\ P^+\ .
\label{dy1}
\end{equation}
{}From energy conservation, we get
\begin{equation}
q^-\ =\ {{\vec q}_{\perp}^2\over P^+\ (\ \beta -\Delta\ )}\ .
\label{dy2}
\end{equation}
Then we have the relation
\begin{equation}
q^2\ =\ q^+q^-\ -\ {\vec q}_{\perp}^2\ =\
{\Delta\over \beta -\Delta}\ {\vec q}_{\perp}^2 \ ,
\label{dy3}
\end{equation}
where
\begin{equation}
\beta \ \ge \ \Delta \ .
\label{dy4}
\end{equation}
Further details on the kinematics are given in the Appendix.

The covariant expression for the four one-loop amplitudes of
diagram Fig. 1(b) is:
\begin{eqnarray}
&&{\cal A}^{\rm one-loop}(I)
\label{fa1s}\\
&=&ig\ e_1\ (\ -e_1\ e_2\ )\ \int {d^4k\over (2\pi)^4}
\nonumber\\
&&\times
{{\rm {\cal N}}(I)\over
(k^2-m^2+i\epsilon )\ ((-k+q)^2-m^2+i\epsilon )
((k-r)^2-\lambda_g^2+i\epsilon )((k-P)^2-\lambda^2+i\epsilon
)}
\nonumber\\
&=&-ig\ e_1\ (\ -e_1\ e_2\ )\ \int {d^2{\vec{k}}_{\perp}\over 2(2\pi)^4}
\int P^+dx\ {{\rm {\cal N}}(I)\over P^{+4}\ x\ (\beta -x)\ (x-\Delta )\ (1-x)}\
\nonumber\\
&&\times
\int dk^- {1\over
\left(k^--{(m^2+{\vec{k}}_{\perp}^2)-i\epsilon\over xP^+}\right)
\left((-k^-+q^-)-{(m^2+(-{\vec{k}}_{\perp}+{\vec{q}}_{\perp})^2)
-i\epsilon\over (\beta -x)P^+}\right)}
\nonumber\\
&&\times
{1\over
\left((k^--r^-)-{(\lambda_g^2+({\vec{k}}_{\perp}-{\vec{r}}_{\perp})^2)
-i\epsilon\over (x-\Delta )P^+}\right)
\left((k^--P^-)+{(\lambda^2+{\vec{k}}_{\perp}^2)-i\epsilon\over
(1-x)P^+}\right)},\
\nonumber
\end{eqnarray}
where we used $k^+=xP^+.$  The numerators ${\rm {\cal N}}(I)$ are given by
\begin{eqnarray}
{\rm {\cal N}}(\Uparrow \to \downarrow)&=&
N\ \ (M+{m\over x})
\label{s9s}\\
{\rm {\cal N}}(\Downarrow \to \downarrow)&=&
N\ \ ({+k^1-ik^2\over x})
\label{s10s}\\
{\rm {\cal N}}(\Uparrow \to \uparrow)&=&
N\ \ ({-k^1-ik^2\over x})
\label{s11s}\\
{\rm {\cal N}}
(\Downarrow \to \uparrow)&=&
N\ \ (M+{m\over x})
\ ,
\label{s12s}
\end{eqnarray}
where
\begin{equation}
N\ =\ 2P^+{\sqrt{\beta -\Delta}}\ \ x\ \, q^- \
\Bigl( -P^+[(1-x)+(1-\Delta )]\Bigr)\ ,
\label{s12sa}
\end{equation}
and $q^-={{\vec q}_{\perp}^2\over P^+( \beta -\Delta )}\ $ as given in
(\ref{dy2}).  For
the [current]-[gauge boson propagator]-[current] factor, in Feynman
gauge only the $-g^{+-}$ term of the gauge boson propagator
$-g^{\mu\nu}$ contributes in the Bjorken limit, and it provides a
factor proportional to $q^-$ in the numerator which cancels the
$q^-$ in the denominator which provides the imaginary part.
Therefore the result scales in the Bjorken limit.

The integration over $k^-$ in (\ref{fa1s}) does not give zero only if
$0 < x < 1$.
We first consider the region $\Delta < x < 1$.
\begin{eqnarray}
&&{\cal A}^{\rm one-loop}(I)
\label{fa1s2}\\
&=&-ig\ e_1\ (\ -e_1\ e_2\ )\ \times\ (2\pi i)\ \int
{d^2{\vec{k}}_{\perp}\over 2(2\pi)^4}
\int P^+dx\ {{\rm {\cal N}}(I)\over P^{+4}\ x\ (\beta -x)\ (x-\Delta )\ (1-x)}\
\nonumber\\
&&\times
{1\over
\left(P^--{(\lambda^2+{\vec{k}}_{\perp}^2)-i\epsilon\over
(1-x)P^+}
-{(m^2+{\vec{k}}_{\perp}^2)-i\epsilon\over xP^+}\right)
\left(-P^-+{(\lambda^2+{\vec{k}}_{\perp}^2)-i\epsilon\over
(1-x)P^+}
+q^--{(m^2+(-{\vec{k}}_{\perp}+{\vec{q}}_{\perp})^2)
-i\epsilon\over (\beta -x)P^+}\right)}
\nonumber\\
&&\times
{1\over
\left(P^--{(\lambda^2+{\vec{k}}_{\perp}^2)-i\epsilon\over
(1-x)P^+}
-r^--{(\lambda_g^2+({\vec{k}}_{\perp}-{\vec{r}}_{\perp})^2)
-i\epsilon\over (x-\Delta )P^+}\right)
},\
\nonumber
\end{eqnarray} where we used $k^+=xP^+$.  The result is identical
to that obtained from light-cone time-ordered perturbation theory.

The phases $\chi_i$ needed for single-spin asymmetries come from the
imaginary part of (\ref{fa1s2}),
which arises from the potentially real intermediate state allowed before
the rescattering.
The imaginary part of the propagator (light-cone energy denominator) gives
\begin{eqnarray}
-i\pi &\delta& \left( -P^-+{(\lambda^2+{\vec{k}}_{\perp}^2)\over
(1-x)P^+} +q^--{(m^2+(-{\vec{k}}_{\perp}+{\vec{q}}_{\perp})^2)
\over (\beta -x)P^+}\right)\nonumber \\
& =& \ -i\pi\ {1\over P^+}\ {(\beta -\Delta)^2\over {\vec q}_{\perp}^2}\
\delta (x\ -\ \Delta \ -\ {\bar{\delta}})\ , \label{sa1}
\end{eqnarray}
where
\begin{equation}
{\bar{\delta}}\ =\ 2\ (\beta -\Delta )\ {{\vec q}_{\perp}\cdot
({\vec k}_{\perp}-{\vec r}_{\perp})\over {\vec q}_{\perp}^2} \ .
\label{sa2}
\end{equation}

Since the exchanged momentum $\bar \delta P^+$ is small, the
light-cone energy denominator corresponding to the
gauge boson propagator is dominated by the $({\vec k}_{\perp}-{\vec
r}_{\perp})^2 + \lambda^2_g \over (x-\Delta)$ term.  This gets
multiplied by $(x-\Delta)$, so only $({\vec k}_{\perp}-{\vec
r}_{\perp})^2 + \lambda^2_g$ appears in the propagator,
independent of whether the photon is absorbed or emitted.  The
contribution from the region $0 < x < \Delta$ thus compliments the
contribution from the region $\Delta < x < 1$.

We can integrate (\ref{fa1s2}) over the transverse momentum using
a Feynman parame\-tri\-zation to obtain the one-loop terms in
(\ref{s1}) to (\ref{s4}).
\begin{eqnarray}
g_1&=&\ \int_0^1d\alpha\ {1\over \alpha (1-\alpha){\vec r}_{\perp}^2
+\alpha \lambda_g^2 +(1-\alpha)\Delta (1-\Delta)
(-M^2+{m^2\over\Delta}+{\lambda^2\over 1-\Delta})}
\label{s7}\\
g_2&=&\ \int_0^1d\alpha\ {\alpha\over \alpha (1-\alpha){\vec
r}_{\perp}^2 +\alpha \lambda_g^2 +(1-\alpha)\Delta (1-\Delta)
(-M^2+{m^2\over\Delta}+{\lambda^2\over 1-\Delta})} \ . \label{s8}
\end{eqnarray}

We define:
\begin{eqnarray}
{\cal P}_z&=&{\cal C}^{-1}\ \Bigl(\ {|A(\Uparrow \to
\uparrow)|^2-|A(\Downarrow \to \uparrow)|^2 +|A(\Uparrow \to
\downarrow)|^2-|A(\Downarrow \to \downarrow)|^2} \ \Bigr)
\label{s16}\\ {\cal P}_x&=&{\cal C}^{-1}\ \Bigl(\
 (\ A(\Uparrow \to \uparrow)^*A(\Downarrow \to \uparrow)\ +\
A(\Uparrow \to \uparrow)A(\Downarrow \to \uparrow)^*\ )
\label{s17}\\ &&\ \ \ +\
 (\ A(\Uparrow \to \downarrow)^*A(\Downarrow \to \downarrow)\ +\
A(\Uparrow \to \downarrow)A(\Downarrow \to \downarrow)^*\ ) \
\Bigr) \nonumber\\ {\cal P}_y&=&{\cal C}^{-1}\ \Bigl(\
 i\ (\ A(\Uparrow \to \uparrow)^*A(\Downarrow \to \uparrow)\ -\
A(\Uparrow \to \uparrow)A(\Downarrow \to \uparrow)^*\ )
\label{s18}\\ &&\ \ \ +\
 i\ (\ A(\Uparrow \to \downarrow)^*A(\Downarrow \to \downarrow)\ -\
A(\Uparrow \to \downarrow)A(\Downarrow \to \downarrow)^*\ ) \
\Bigr) \ , \nonumber
\end{eqnarray}
where the normalization from the unpolarized cross section is
\begin{equation}
{\cal C}=|A(\Uparrow \to \uparrow)|^2+|A(\Downarrow \to
\uparrow)|^2 +|A(\Uparrow \to \downarrow)|^2+|A(\Downarrow \to
\downarrow)|^2 \ . \label{s19}
\end{equation}

We can assume for convenience that the initial-state interactions
generate a phase when exponentiated, as in the Coulomb phase
analysis of QED.  The rescattering phases $e^{i\chi_i}$ $(i =
1,2)$ with $\chi_i = {\rm tan}^{-1}({e_1 e_2 \over 8 \pi
}{g_i\over h})$ are thus distinct for the spin-parallel and
spin-antiparallel amplitudes.  The difference in phase arises from
the orbital angular momentum $k_\perp$ factor in the spin-flip
amplitude, which after integration gives the extra factor of the
Feynman parameter $\alpha$ in the numerator of $g_2$ compared to
$g_1$, as we can see in (\ref{s7}) and (\ref{s8}). Notice that the
phases $\chi_i$ are each infrared divergent for zero gauge boson
mass $\lambda_g \to 0$, as is characteristic of Coulomb phases.
However, the difference $\chi_1 - \chi_2$ which contributes to the
single-spin asymmetry is infrared finite. We have verified that
the Feynman gauge result is also obtained in the light-cone gauge
using the principal value prescription. The small numerator
coupling of the light-cone gauge boson is compensated by the small
value for the exchanged $l^+ = \bar \delta P^+$ momentum.

The virtual photon  and produced hadron define the  production
plane which we will take as the ${\hat z}-{\hat x}$ plane. {}From
Eqs. (\ref{s1})-(\ref{s4}) and  (\ref{s18}), the azimuthal
single-spin asymmetry transverse to the production plane is given
by
\begin{eqnarray}
{\cal P}_y &=& -\ {e_1e_2\over 8\pi} \ {2\ \Bigl(\ \Delta\, M+m\
\Bigr)\ r^1\over \Big[\ \Bigl( \ \Delta\, M+m\ \Bigr)^2\ +\ {\vec
r}_{\perp}^2\ \Big]}\ \Big[\ {\vec r}_{\perp}^2+\Delta
(1-\Delta)(-M^2+{m^2\over\Delta} +{\lambda^2\over 1-\Delta})\
\Big] \nonumber\\ &\times& \ {1\over {\vec r}_{\perp}^2}\ {\rm
ln}{{\vec r}_{\perp}^2 +\Delta
(1-\Delta)(-M^2+{m^2\over\Delta}+{\lambda^2\over 1-\Delta})\over
\Delta (1-\Delta)(-M^2+{m^2\over\Delta}+{\lambda^2\over
1-\Delta})}\ ,\label{sa2b1}
\end{eqnarray}
in agreement with the crossing properties described in section 2.
The linear factor of $r^1 = r^x$ reflects the fact that the
single-spin asymmetry is proportional to ${\vec S}_p \cdot \vec q
\times \vec r$ since $\vec q  \sim - \hat z |\vec q|$ and ${\vec
S}_p = \pm \hat y.$  The kinematics are given in more detail in
the Appendix.  The prediction for ${\cal P}_y $ as a function of
$\Delta$ and $r_\perp$ is identical but with opposite sign to that
illustrated in Fig. 4 of Ref.~\cite{Brodsky:2002cx}.

Our analysis can be generalized to the corresponding calculation
in QCD.  The initial-state interaction from gluon exchange has the
strength ${e_1 e_2\over 4 \pi} \to C_F \alpha_s(\mu^2).$ The scale
of $\alpha_s$ in the ${\overline{MS}}$ scheme can be identified
with the momentum transfer carried by the gluon $\mu^2= e^{-5/3}
({\vec k}_{\perp}-{\vec r}_{\perp})^2$ \cite{Brodsky:1995ds}. The
matrix elements coupling the proton to its constituents will have
the same numerator structure as the perturbative model since they
are determined by orbital angular momentum constraints.  The
strengths of the proton matrix elements can be normalized quark by
quark according to their contributions to the target nucleon's
anomalous magnetic moment weighted by the quark charge squared. In
QCD, $r_\perp$ is the magnitude of the momentum of the current
quark jet relative to the virtual photon direction. Notice that
for large $r_\perp$, ${\cal P}_y$ decreases as $
\alpha_s(r^2_\perp) x_{bj} M r_\perp \ln r^2_\perp \over
{r}_{\perp}^2 $.  The physical proton mass $M$ appears since it is
present in the ratio of the $L^z = 1$ and $L^z = 0$ matrix
elements.  This form is expected to be essentially universal.

\section{Summary}

We have shown that the same physical mechanism which produces a
leading-twist single-spin asymmetry in semi-inclusive DIS, also
leads to a leading twist single-spin asymmetry in the Drell-Yan
process. The initial-state interaction between the annihilating
antiquark with the spectator of the target produces the required
phase correlation.  The equality in magnitude, but opposite sign,
of the single-spin asymmetries in semi-inclusive DIS and the
corresponding Drell-Yan processes is an important check of our
mechanism.

It has been conventional to assume that the effects of initial-
and final-state interactions are always power-law suppressed for
hard processes in QCD.   In fact, this is not in general correct,
as can be seen from our analyses of leading-twist single-spin
asymmetries in the Drell-Yan process and semi-inclusive deep
inelastic scattering.  The initial- and final-state interactions
which survive in the scaling limit occur in light-cone time $\tau
= {\cal O}(1/Q)$ immediately before or after the hard subprocess.
Other initial- and final-state interactions, such as those between
the spectator of the incident hadron and the spectator of the
target hadron in the DY process, take place over long time scales,
and they only provide inconsequential unitary phase corrections to
the process. This is in accord with our intuition that
interactions which occur at distant times cannot affect the
primary reaction.

A natural framework for the wavefunctions which appear in the SSA
calculations is the light-front Fock expansion
\cite{Lepage:1980fj,Brodsky:1989pv}. In principle, the light-front
wavefunctions for hadrons can be obtained by solving for the
eigen-solutions of the light-front QCD Hamiltonian. Such
wavefunctions are real and include all interactions up to a given
light-front time.  The final-state gluon-exchange corrections
which provide the SSA for semi-inclusive DIS occurs immediately
after the virtual photon strikes the active quark. Such
interactions are not included in the light-front wavefunctions,
just as Coulomb final-state interactions are not included in the
Schr\"odinger bound state wavefunctions in QED.
Collins~\cite{Collins} has argued that since the relevant
rescattering interactions of the struck quark occur very close in
light-cone time to the hard interaction, one can augment the
light-front wavefunctions by a Wilson line factor which
incorporates the effects of the final-state interactions in
semi-inclusive DIS. However, such augmented wavefunctions are not
universal and process independent; for example, in the case of the
DY process, an incoming Wilson line of opposite phase must be
used.

Our formalism can be adopted to single-spin asymmetries in more
general hard inclusive reactions, such as $\bar p p^\uparrow \to
\pi X,$ where the pion is detected at high transverse
momentum~\cite{E70496,lambda}. In such cases one must identify the
hard quark-gluon subprocess and analyze a set of gluon exchange
corrections which connect the spectators of the polarized hadron
with the active quarks and gluons of the hard subprocess. An
example of a final-state interaction which can cause a single
asymmetry in  $\bar p p^\uparrow \to \pi X$ is shown in Fig. 2.
However, this type of final-state interaction cannot be readily
identified as an augmented target wavefunction. It is also clear
from our analyses that there are potentially important corrections
to the hard quark propagator in hard exclusive subprocesses such
as deeply virtual Compton scattering or exclusive meson
electroproduction.  These rescattering interactions of the
propagating quark can provide new single-spin observables and will
correct analyses based on the handbag approximation.

\begin{figure}
\centering
\includegraphics{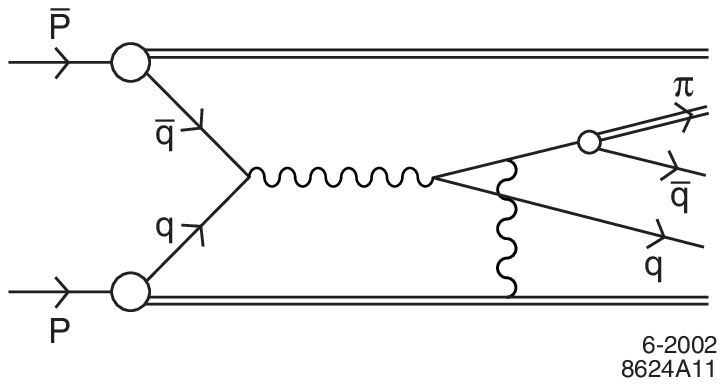}
\caption[*]{An example of a final-state interaction which can
cause a single asymmetry in  $\bar p p^\uparrow \to \pi X$.}
\label{fig:2}
\end{figure}

It should be emphasized that the same overlap of light-front
wavefunctions  with $\Delta L^z = 1$  which gives single-spin
asymmetries also yields the Pauli form factor $F_2(t)$ and the
generalized parton distribution $E(x,\zeta,t)$ entering deeply
virtual Compton
scattering~\cite{Muller:1998fv,Ji:1996ek,Radyushkin:1997ki,
Diehl:2000xz,Brodsky:2000xy}. Each quark of the target
wavefunction appears additively, weighted linearly by the quark
charge in the case of the Pauli form factor and weighted
quadratically in the case of deep inelastic scattering, the
Drell-Yan reaction and deeply virtual Compton scattering.

The empirical study of single-spin asymmetries in hard inclusive
and exclusive processes thus provides a new window to the
investigation of hadron spin, angular momentum, and flavor
structure of hadrons.

\bigskip

\noindent {\large \bf Acknowledgments}

We thank Harut Avakian, John Collins, Paul Hoyer, Xiang-dong Ji,
and Stephane Peigne for helpful conversations.

\section*{Appendix: Kinematics }

{}From Fig. 1 we have [$ a =(a^+,a^-,{\vec a}_{\perp})$]
\begin{eqnarray}
P&=& \Bigl( P^+\ ,\ {M^2\over P^+}\ ,\ {\vec 0}_{\perp}\Bigr)
\label{ap1}\\
P'&=& \Bigl( (1-\Delta)P^+\ ,\ {\lambda^2+{\vec r}_{\perp}^2\over
(1-\Delta)P^+} \ ,\ -{\vec r}_{\perp}\Bigr)
\label{ap2}\\
q&=& \Bigl( \beta P^+\ ,\ {q^2+{\vec
q}_{\perp}^2\over \beta P^+}\ ,\ {\vec q}_{\perp}\Bigr)   \label{ap3}\\
P_{\bar{q}}&=&  \Bigl( (\beta-\Delta) P^+\ ,\ {m^2+({\vec
q}_{\perp}-{\vec r}_{\perp})^2\over (\beta -\Delta)P^+}\ ,\ {\vec
q}_{\perp}-{\vec r}_{\perp}\Bigr) \ , \label{ap4}
\end{eqnarray}
where
\begin{equation}
s - m^2 - M^2  = 2 P_{\bar q} \cdot P = {m^2+({\vec
q}_{\perp}-{\vec r}_{\perp})^2\over (\beta -\Delta)} + M^2 (\beta
- \Delta ) \  . \label{ap4aa}
\end{equation}

Energy conservation gives
\begin{equation}
{m^2+({\vec q}_{\perp}-{\vec r}_{\perp})^2\over (\beta -\Delta)}
+{M^2 }= {q^2+{\vec q}_{\perp}^2\over \beta } +{\lambda^2+{\vec
r}_{\perp}^2\over (1-\Delta)}\ . \label{ap5}
\end{equation}
Here $s$ and $q^2 $ are large. It is convenient to work in a frame
where $\beta - \Delta = {\cal O}(1)$ and ${\vec q}_{\perp}^2$ is
large so that $s \simeq {{\vec q}^2_\perp\over \beta - \Delta}. $
In such a frame, $\vec q  \sim - \hat z |\vec q|.$  Then Eq.
(\ref{ap5}) gives
\begin{equation}
{\vec q}_{\perp}^2={\beta -\Delta\over \Delta}q^2\ . \label{ap6}
\end{equation}

{}From (\ref{ap1}), (\ref{ap3}) and (\ref{ap6}) we have
\begin{equation}
2P\cdot q={ {\vec q}_{\perp}^2\over \beta-\Delta}+\beta M^2\
\simeq {q^2\over \Delta} , \label{ap7}
\end{equation}
and thus to leading twist
\begin{equation}
\Delta \ =\ {q^2\over 2P\cdot q }\ , \label{ap8}
\end{equation}
where $P\cdot q/M = \nu$ is the energy of the lepton pair in the
target rest frame.

{}From (\ref{ap1}) and (\ref{ap2}) we have
\begin{equation}
2P\cdot P'={\lambda^2+{\vec r}_{\perp}^2\over
1-\Delta}+(1-\Delta)M^2\ . \label{ap9}
\end{equation}
Since $t=(P-P')^2$, we have $2P\cdot P'=-t+M^2+\lambda^2$.
Therefore, (\ref{ap9}) gives
\begin{equation}
{\vec r}_{\perp}^2\ =\ (1-\Delta)\Bigl(
-t+M^2+\lambda^2-(1-\Delta)M^2\Bigr) -\lambda^2\ . \label{ap10}
\end{equation}
which relates ${\vec r}_{\perp}^2$ to invariants.

\end{document}